\documentclass[12pt]{iopart}

\usepackage{acronym}
\usepackage[OT2,T1]{fontenc}
\usepackage{gensymb}
\usepackage{wasysym}
\usepackage{graphicx}
\usepackage{multirow} 
\usepackage{color}

\newcommand{\geant}{{\sc{Geant4}}}
\newcommand{\labr}{LaBr$_{3}$:Ce}
\newcommand{\cebr}{CeBr$_{3}$}

\newcommand{\talys}{{\sf{TALYS}}}

\begin{document}

\title[Nuclear level density of ${}^{128}$Te]{Nuclear level density of ${}^{128}$Te from $(\mathrm{p},\mathrm{p}'\gamma)$ scattering and complementary photonuclear data}

\author{P.-A.~Söderström\textsuperscript{1,*}, 
A.~Ku\c{s}o\u{g}lu\textsuperscript{1,2}, 
S.~Aogaki\textsuperscript{1}, 
D.~L.~Balabanski\textsuperscript{1}, 
S.-R.~Ban\textsuperscript{1}, 
R.~Borcea\textsuperscript{3}, 
M.~Brezeanu\textsuperscript{1}, 
S.~Calinescu\textsuperscript{3}, 
C.~Costache\textsuperscript{3}, 
R.~Corbu\textsuperscript{1}, 
M.~Cuciuc\textsuperscript{1}, 
A.~Dhal\textsuperscript{1}, 
I.~Dinescu\textsuperscript{3}, 
N.~M.~Florea\textsuperscript{3}, 
T.~Furuno\textsuperscript{4}, 
A.~Gavrilescu\textsuperscript{1}, 
A.~Gupta\textsuperscript{5}, 
Y.~Honda\textsuperscript{4}, 
J.~Isaak\textsuperscript{5}, 
N.~C.~Jerca\textsuperscript{1}, 
T.~Kawabata\textsuperscript{4}, 
V.~Lelasseux\textsuperscript{1}, 
R.~Lica\textsuperscript{3}, 
C.~Marin\textsuperscript{1}, 
C.~Mihai\textsuperscript{3}, 
S.~Niculae\textsuperscript{1}, 
H.~Pai\textsuperscript{1}, 
I.~P.~P\^{a}rlea\textsuperscript{1}, 
T.~Petruse\textsuperscript{1}, 
A.~Spataru\textsuperscript{1}, 
D.~A.~Testov\textsuperscript{1}, 
D.~Tofan\textsuperscript{3}, 
T.~Tozar\textsuperscript{1}, 
A.~Turturic\u{a}\textsuperscript{3}, 
G.~V.~Turturic\u{a}\textsuperscript{1}, 
S.~Ujeniuc\textsuperscript{3}}

\address{\textsuperscript{1} Extreme Light Infrastructure (ELI-NP), Horia Hulubei National Institute for R\&D in Physics and Nuclear Engineering (IFIN-HH), Str. Reactorului No. 30, 077125 Bucharest--M\u{a}gurele, Romania\\
\textsuperscript{2} Department of Physics, Faculty of Science, Istanbul University, Vezneciler/Fatih, 34134, Istanbul, Turkey\\
\textsuperscript{3} Department of Nuclear Physics, Horia Hulubei National Institute for R\&D in Physics and Nuclear Engineering (IFIN-HH), Str. Reactorului No. 30, 077125 Bucharest--M\u{a}gurele, Romania\\
\textsuperscript{4} Department of Physics, Osaka University, Toyonaka, Osaka 560-0043, Japan\\
\textsuperscript{5} Institute for Nuclear Physics, Department of Physics, Technische Universität Darmstadt, Darmstadt 64289, Germany
}
\ead{par.anders@eli-np.ro}
\vspace{10pt}
\begin{indented}
\item[]\today
\end{indented}

\begin{abstract}
We have extracted the nuclear level density of ${}^{128}$Te from a $(\mathrm{p},\mathrm{p} '\gamma)$ scattering experiment using the large-volume \labr\ and \cebr\ detectors from ELI-NP at the 9~MV Tandem facilities at IFIN-HH. The decay data were normalised using photonuclear data, resulting in nuclear level densities without intrinsic model dependencies from the constant temperature or Fermi gas models. The deduced nuclear level density follows in between the expectations from these two models, but we observe a clear divergence from a microscopic model based on the Skyrme force. 
\end{abstract}

%
%
%
%
%

\section{Introduction}

Statistical nuclear properties, like \acp{gSF} and \acp{NLD}, are important measurable observables determining reaction rates for applications like nuclear reactor technologies and criticality, processing of spent nuclear fuel, and nuclear astrophysics. These observables serve as input to the statistical model that describes the atomic nucleus at energies and temperatures where discrete states start to overlap due to their natural widths and can not be treated as separate energy states anymore. For a recent overview of the current experimental and theoretical status of \acp{NLD} and \ac{gSF}, see Reference~\cite{Wiedeking2024}. Several different traditional methods exist for \ac{NLD} measurements. These include well-established approaches, such as the Oslo method \cite{Guttormsen1987,Guttormsen1996,Schiller2000,Larsen2011}, evaporation spectra of protons \cite{Voinov2019} or neutrons \cite{Roy2021}, or high-energy $(\mathrm{p},\mathrm{p}')$ scattering with magnetic spectrometers \cite{Usman2011}.

Another interesting opportunity for experimental studies of \acp{NLD} and \acp{gSF} is the projected $\gamma$-ray beams at the \ac{ELI-NP} \cite{Filipescu2015,Gales2016,Gales2018,Tanaka2020,Constantin2024}. While charged-particle reactions populate an extensive range of states, electromagnetic probes consisting of narrow bandwidth photon beams for photoexcitation and decay studies have high selectivity in excitation energy, spin, and parity of the ensemble of excited states \cite{Zilges2022}. This has been explored in depth in pioneering work at the \ac{HIgS} facility at the \ac{TUNL}, Duke University, North Carolina \cite{Weller2009}, using the $(\vec{\gamma},\gamma'\gamma'')$ technique \cite{Isaak2019}, meaning inelastic scattering of incoming polarised photons, $\vec{\gamma}$, with a two-step decay as $\gamma'$ and $\gamma''$ to extract \acp{gSF} directly. Due to the properties of the $\gamma$-ray beams, the \acp{gSF} could be explicitly determined for dipole-excited states and, also, for well-defined excitation energies providing the potential for a sensitive test of the Brink-Axel hypothesis \cite{Brink1955,Axel1962}, that the \acp{gSF} are independent of the excitation energy in the statistical region and only depend on the energy difference between the initial and final states.

While the $\gamma$-ray beams at \ac{ELI-NP} are under implementation, a complementary scientific program has been started at the charged particle accelerator facilities at the \ac{IFIN-HH}, in particular at the 9~MV Tandem accelerator facilities. In this programme, we utilise the instrumentation from \ac{ELI-NP}, especially the \ac{ELIGANT} instruments \cite{Camera2016,Soderstrom2022} and other setups in the existing \ac{ELI-NP} and \ac{IFIN-HH} infrastructure. One of the primary campaigns has been utilising the \ac{ROSPHERE} \cite{Bucurescu2016} infrastructure and detection system in combination with the large-volume \labr\ and \cebr\ detectors for a system dedicated to high-energy $\gamma$ rays. This special version of \ac{ROSPHERE} \cite{Aogaki2023} has already succeeded in detailed studies of light nuclei and high-energy resonant states in medium-mass nuclei \cite{Kusoglu2024a,Kusoglu2024b,Kusoglu2024c,Kusoglu2024d,Wieland2024a,Soderstrom2024a,Sakanashi2024}. In addition to these experiments, an experiment dedicated to \acp{NLD} and \acp{gSF} on ${}^{112,114}$Sn was performed in 2023 as a proof-of-concept for measurements of statistical properties with the Oslo method at \ac{IFIN-HH} \cite{SoderstromUnp}, and to follow up a recent systematic study of the statistical properties of the Sn chain \cite{Markova2021,Markova2022,Markova2023,Markova2024,Markova2025}. In 2024, a follow-up of the 2023 experiment was performed at the \ac{IFIN-HH} to verify the consistency between data obtained from charged particle beams and $\gamma$-ray beams for the nucleus ${}^{128}$Te for \acp{gSF} within the context of the results from Reference~\cite{Isaak2019}. While the details of this data set are still under analysis, interesting results were obtained when applying the methodology to \acp{NLD}, which will be reported here.

\section{Experiment\label{sec:exp}}

The experiment was conducted at the \ac{IFIN-HH} 9~MV Tandem facilities in M\u{a}gurele, Romania, in spring 2024. The experimental setup consisted of using a combination of the \ac{ELIGANT-GN} large-volume \labr, and \cebr\ detectors \cite{Camera2016,Soderstrom2022} mounted in the mechanical frame and inside the \ac{BGO} shields of the \ac{ROSPHERE} \cite{Bucurescu2016}, a configuration that has been previously reported in detail in Reference~\cite{Aogaki2023}. The total angular coverage of the scintillators in this configuration was 11.95\% of the solid angle. In addition to the $\gamma$-ray detectors, a charged-particle detector array consisting of silicon-strip detectors of the type Micron S7 with thicknesses of 65~$\mu$m and 1000~$\mu$m, respectively, was placed in a $\Delta E - E$ configuration in the backward direction. These detectors were placed at a distance of 28~mm from the target for the thin detector and 44~mm from the target for the thick detector, providing an angular coverage of 122$^{\circ}$-136$^{\circ}$ in the overlap region relative to the target.

The $\gamma$-ray detection part of the setup consisted of 20 \labr\ and \cebr\ detectors from \ac{ELI-NP} optimised for high-energy $\gamma$-rays and four \ac{HPGe} detectors for high-resolution spectroscopy of low-energy transitions to quantify background and verify the spin distributions of the excited states for different excitation energies. For calibration of the $\gamma$-ray detectors simple radioactive sources of ${}^{137}$Cs, ${}^{60}$Co, and ${}^{56}$Co were used for low energies and time alignments. In the high-energy region composite sources consisting of \ac{PuBe} \cite{Soderstrom2021} inside of a sphere filled with a nickel-paraffin mixture \cite{Soderstrom2023a} were used, providing two energy calibration points at 4.44~MeV and 9~MeV. The silicon detectors were aligned using a standard three-$\alpha$ source consisting of ${}^{239}$Pu, ${}^{241}$Am, and ${}^{244}$Cm, as well as from elastic scattering of the protons using the in-beam data.

The target was 1.7~mg/cm$^{2}$ and evaporated on a 30~$\mu$g/cm$^{2}$ backing of diamond-like carbon. The isotopic purity of the target was 98.2(2)\% ${}^{128}$Te with the largest impurity contribution from ${}^{130}$Te (1.25\%), ${}^{126}$Te (0.37\%), and ${}^{125}$Te (0.11\%) while other isotopes of Te contributed with $<0.1$\% each. On this target, a proton beam with an energy of 14~MeV and a typical beam current of $1.5$~nA impinged, and the data were collected for 70~h. The limiting factor in these measurements was the count rate in the \ac{BGO} detectors. Thus, they were disabled in the \ac{DAQ}. The \ac{DAQ} system was fully digital with the \labr\ and \cebr\ detectors read out using CAEN V1730 digitisers running \ac{DPP-PSD} firmware, and the silicon detectors as well as the \ac{HPGe} detectors read out by CAEN V1725 digitisers running \ac{DPP-PHA} firmware.

\section{Analysis\label{sec:analysis}}

\begin{figure*}[ht!]
\begin{center}
\includegraphics[width=0.32\textwidth]{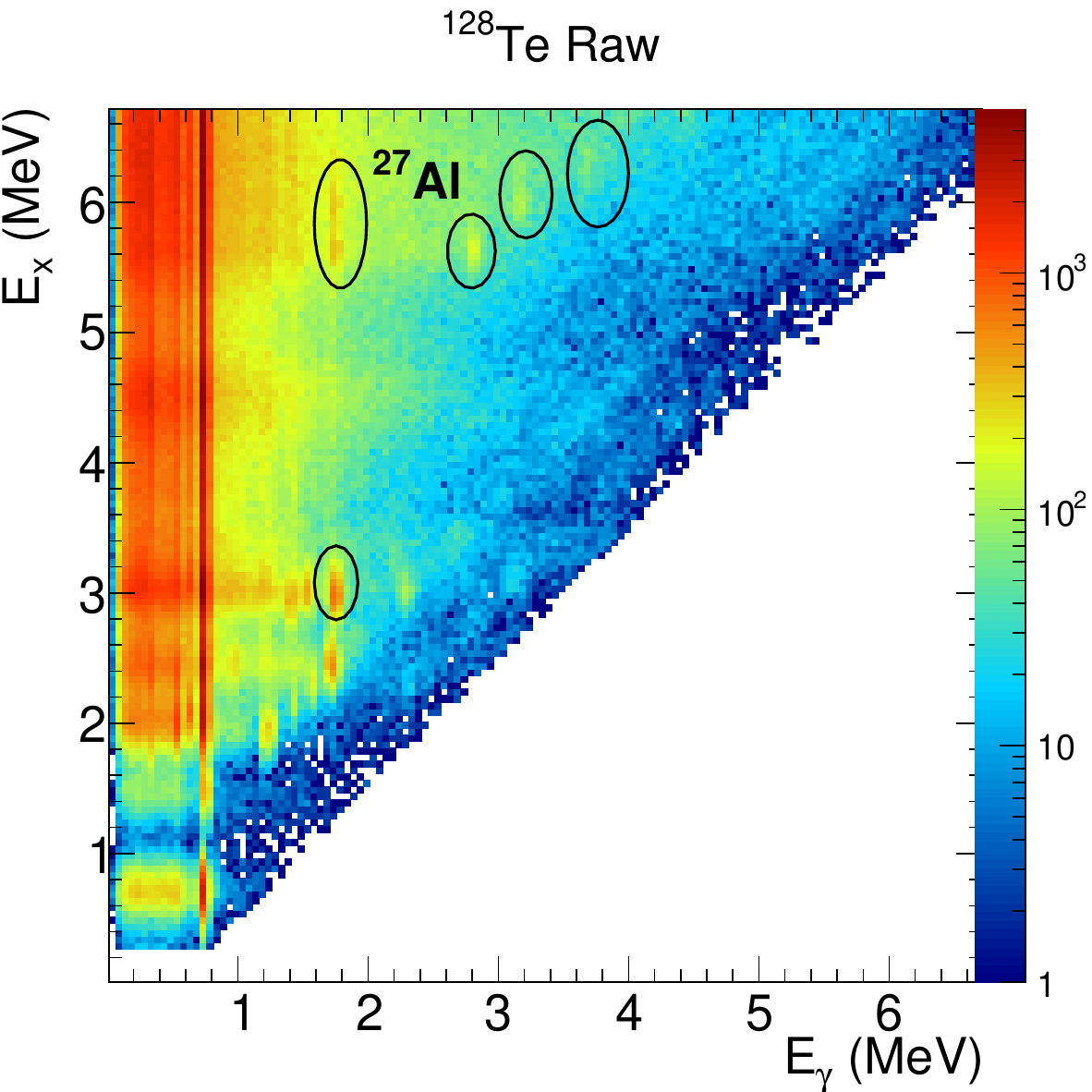}
\includegraphics[width=0.32\textwidth]{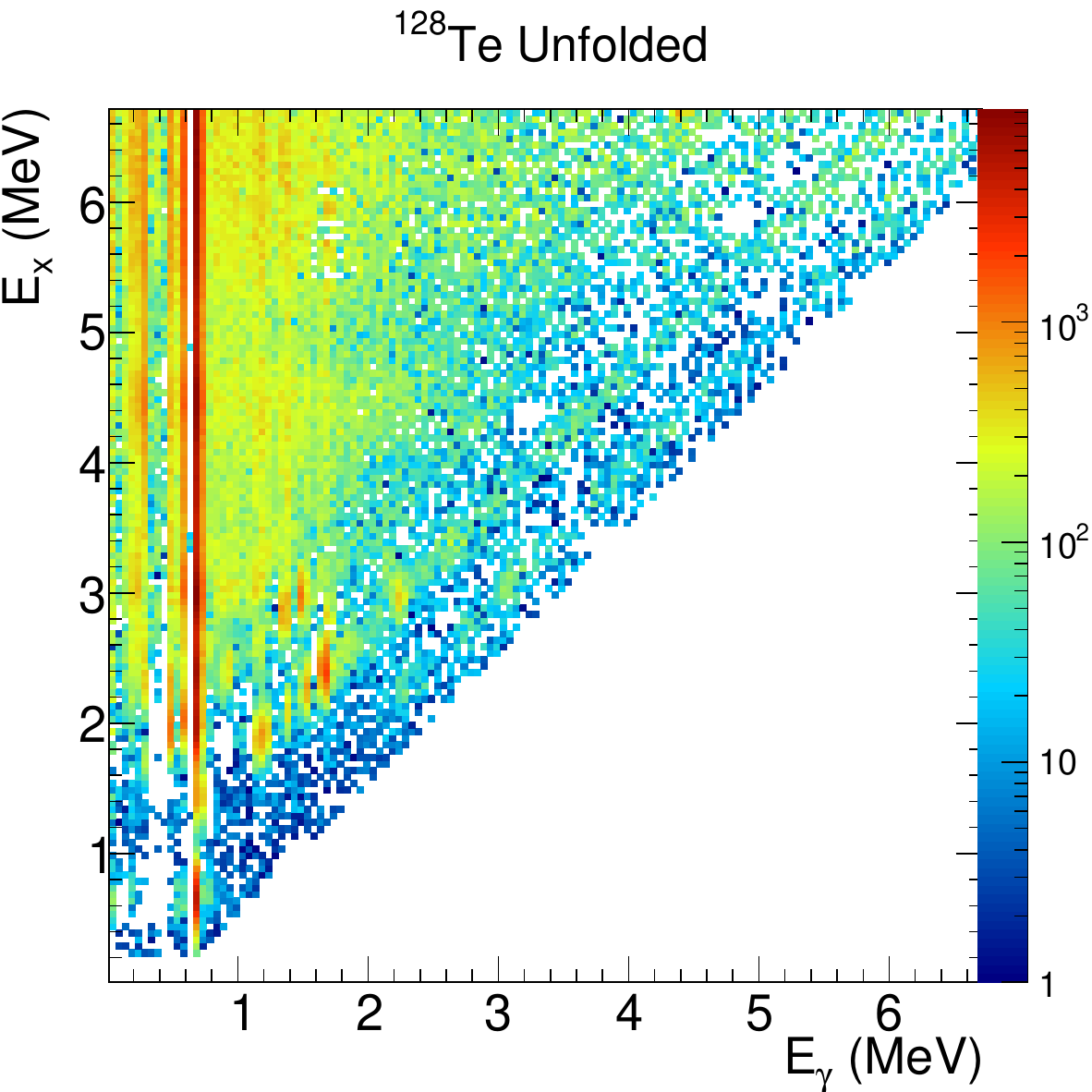}
\includegraphics[width=0.32\textwidth]{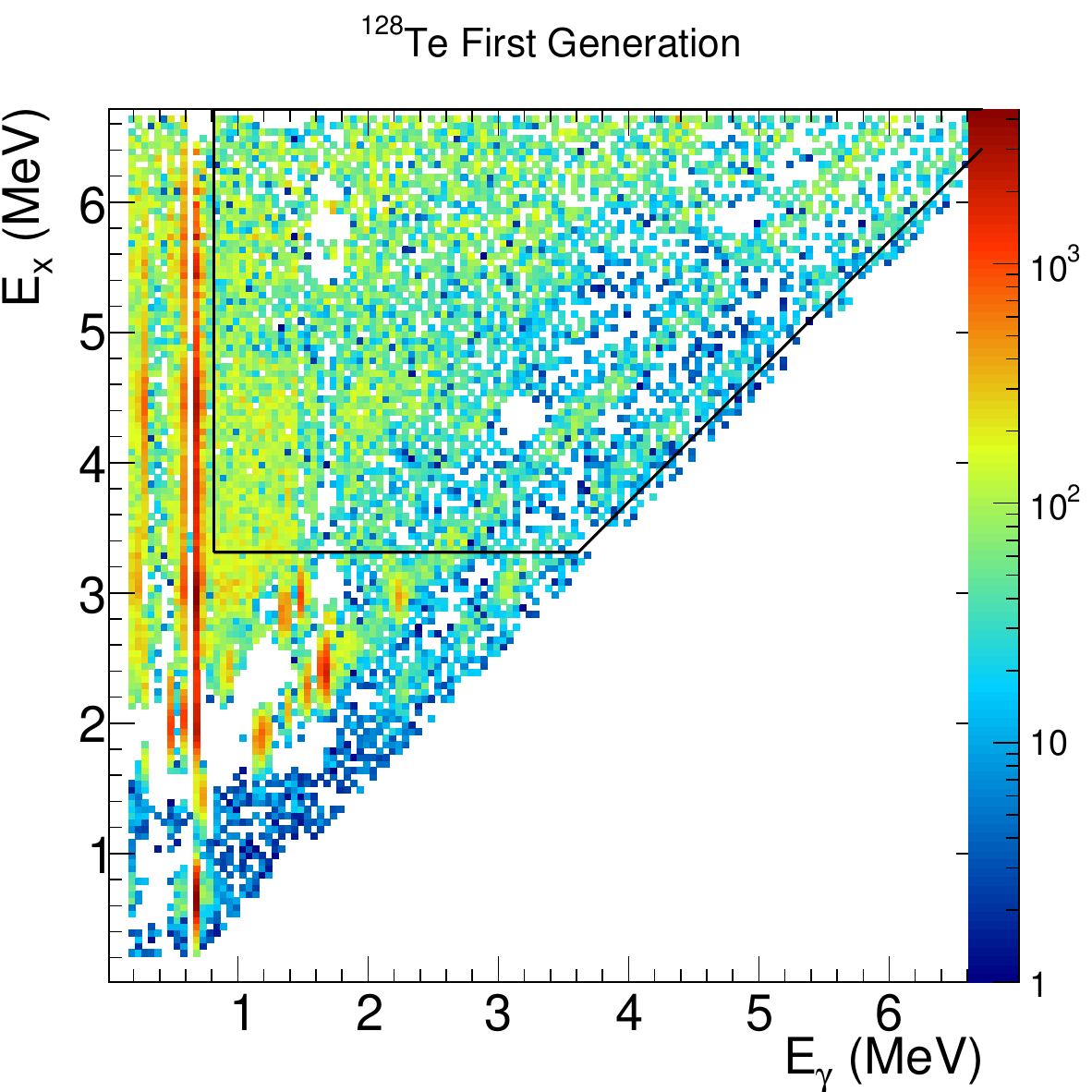}
\end{center}
\caption{Raw (left), unfolded (middle) and first generation (right) matrices for ${}^{128}$Te. The area selected for further analysis of the first-generation matrices is shown as a black outline.}\label{fig:matrices}
\end{figure*}

The \labr, \cebr, and silicon detectors were calibrated in energy and time-aligned using the sources described in Section~\ref{sec:exp}. After calibration, the $\Delta E - E$ telescopes selected events corresponding to scattered protons. The angle of each scattered proton, $\theta_{\mathrm{p}}$, was determined from the $\Delta E$ layer, and the energy of the proton, $E_{\mathrm{p}}$, was measured as the sum $E_{\mathrm{p}} = \Delta E + E$. Given $\theta_{\mathrm{p}}$, the excitation energy of the nucleus, $E_{x}$, was calculated from the energy of an elastically scattered proton, $E_{\mathrm{el.}}(\theta_{\mathrm{p}})$, as $E_{x}=E_{\mathrm{el.}}(\theta_{\mathrm{p}})-E_{\mathrm{p}}$. The $\gamma$-ray spectra from the \labr\ and \cebr\ detectors were then extracted for each value of $E_{x}$ and arranged in a two-dimensional matrix as shown in Figure~\ref{fig:matrices}.
Some discrete peaks identified as background from ${}^{27}$Al have been highlighted and subtracted in the unfolding step, discussed later in this text. Note that these $\gamma$-ray spectra also contain the response function of the \labr\ and \cebr\ detectors and, as we are working in the quasi-continuum region, this response needs to be unfolded from the spectra. For this purpose, the iterative unfolding procedure \cite{Guttormsen1996,Soderstrom2019b,SoderstromUnp} was applied using the simulated detector response from \geant\ \cite{Agostinelli2003} implemented in an in-house developed software, \ac{GROOT} \cite{Lattuada2017}, as shown in Figure~\ref{fig:matrices}. As the decay probability of an excited state only depends on the \ac{NLD} at the final energy, $E_{\mathrm{f}}$ of the first transition and the value of the \ac{gSF} for the energy difference $E_{\gamma} = E_{\mathrm{x}}-E_{\mathrm{f}}$, all contributions from cascading transitions were subtracted using the first-generation methodology described in detail in References~\cite{Guttormsen1987,Schiller2000,Larsen2011}, providing the so-called first generation matrix shown in Figure~\ref{fig:matrices}.

\begin{figure*}[ht!]
\begin{center}
\includegraphics[width=0.49\textwidth]{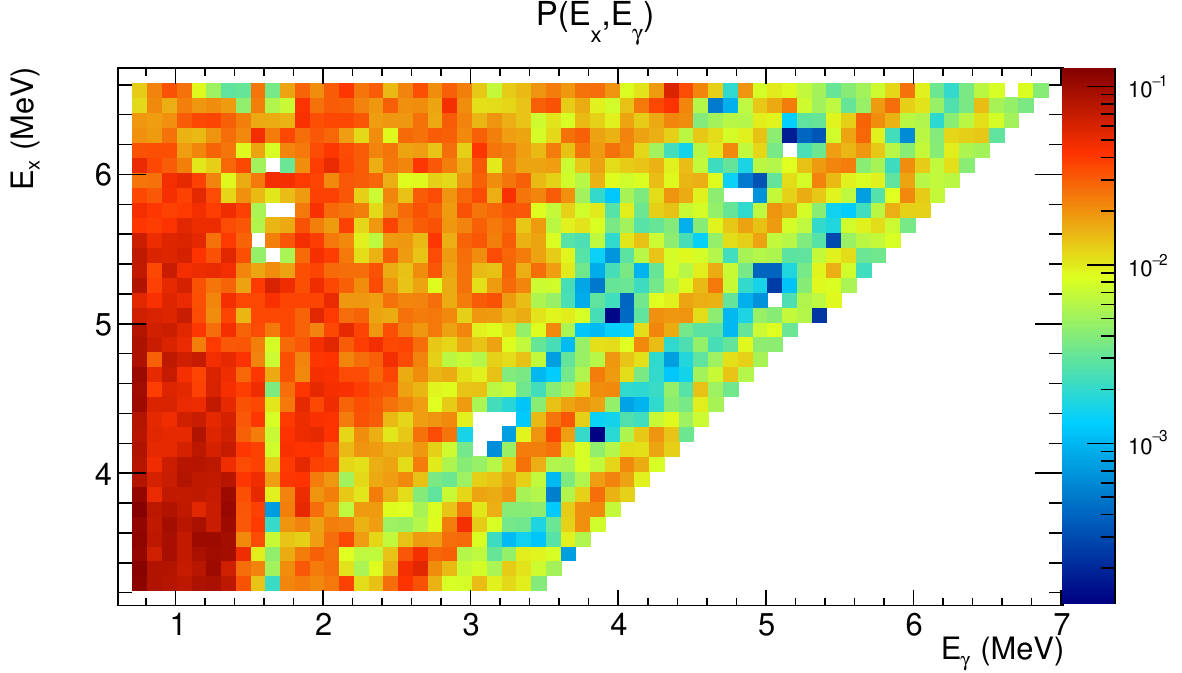}
\includegraphics[width=0.49\textwidth]{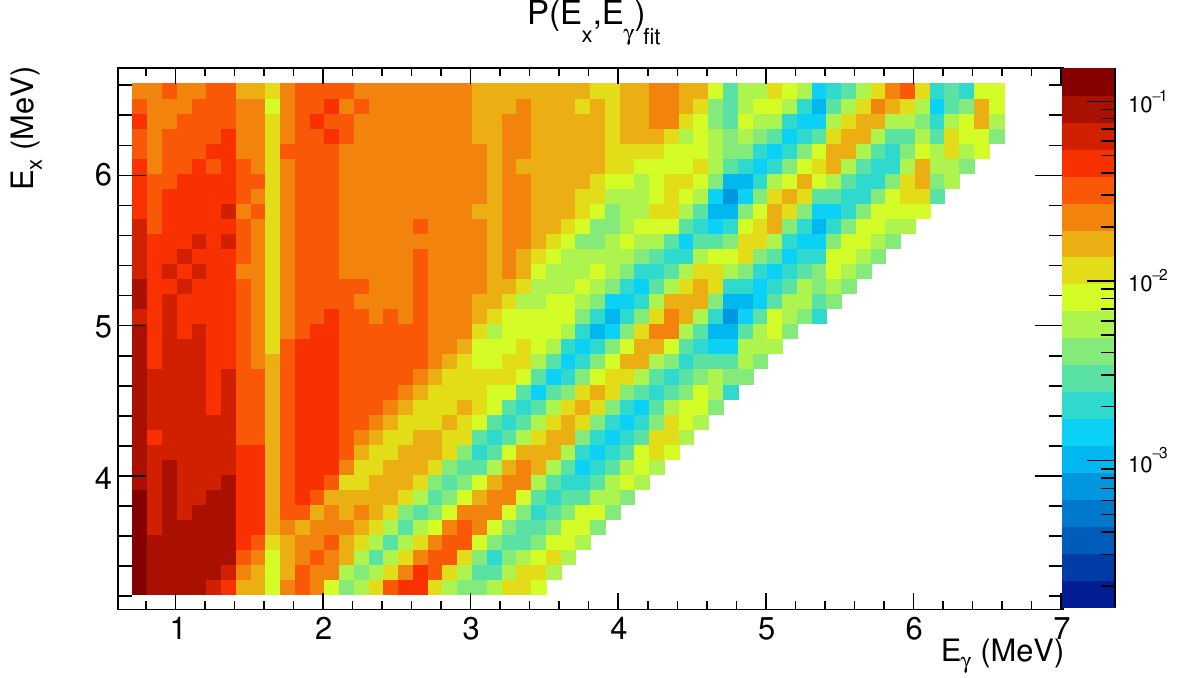}\\
\includegraphics[width=0.49\textwidth]{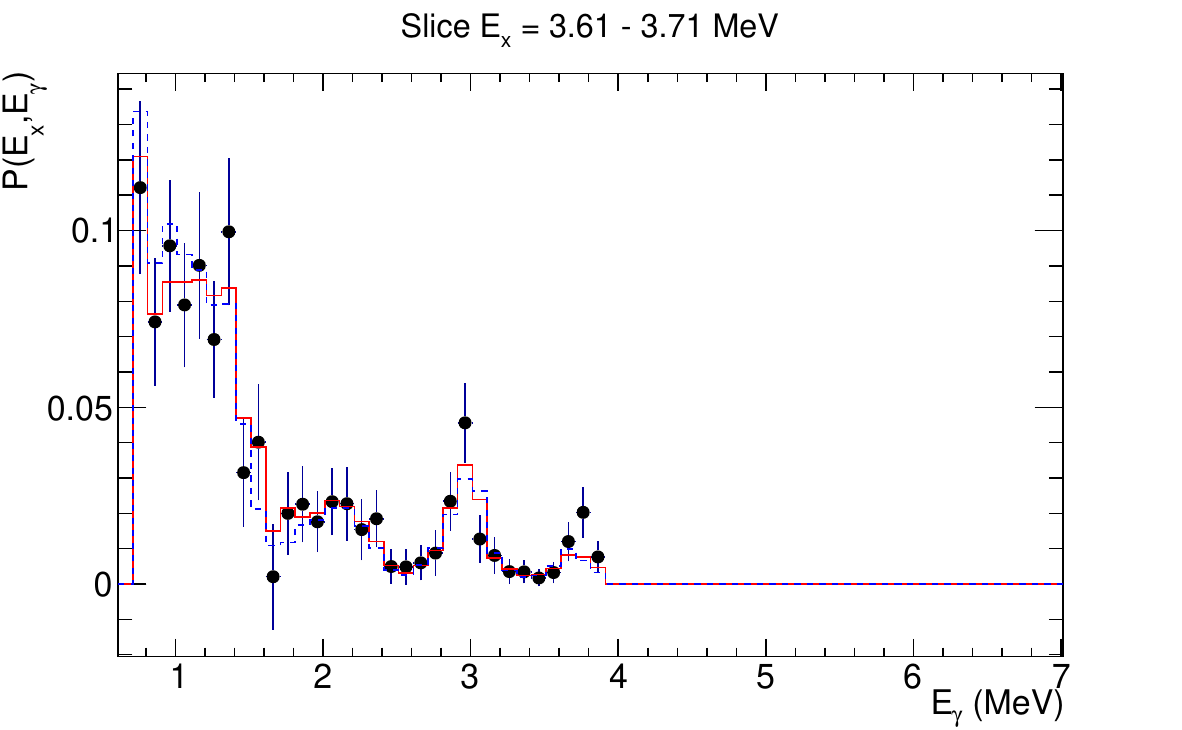}
\includegraphics[width=0.49\textwidth]{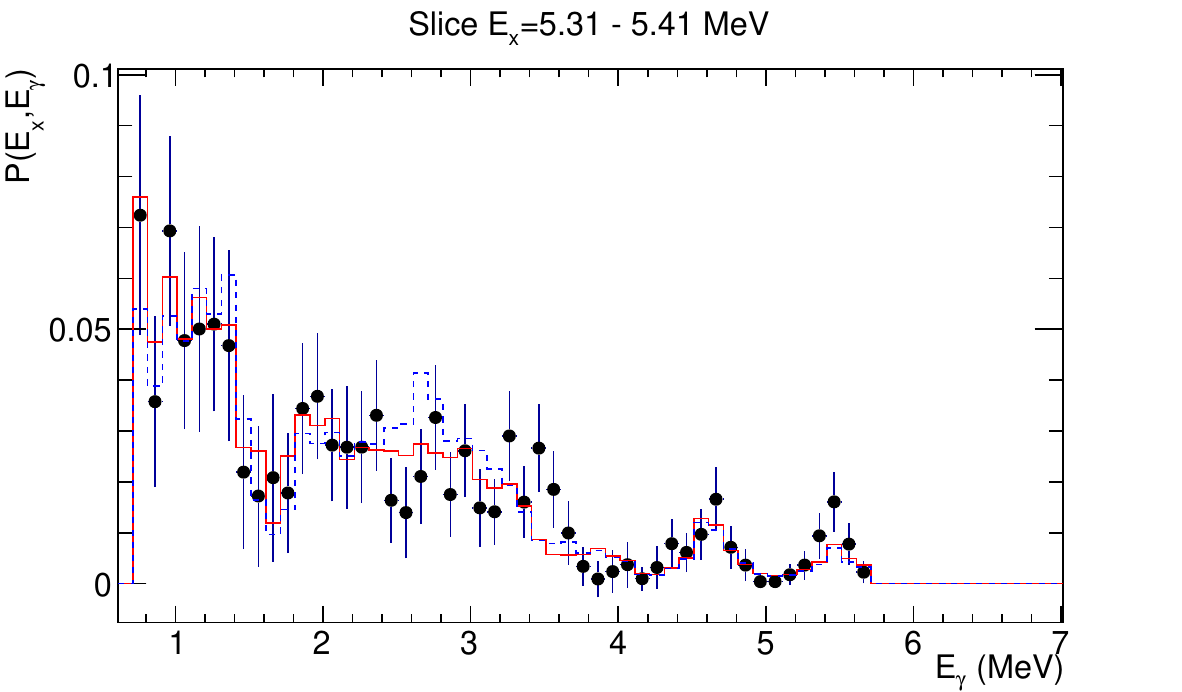}
\end{center}
\caption{Experimental (top left) and best fit (top right) of the probability matrix. The bottom panels show two projections of the best fit (red line) compared with the experimental data (black circles) in the energy ranges 3.61-3.71~MeV and 5.31-5.41~MeV. For comparison, the best fit of the data without the aluminium contamination subtracted (blue dotted line) is also shown. The error bars include both statistical uncertainties and systematic uncertainties evaluated for the reduced $\chi^{2}/\mathrm{NDF} = 1$.}\label{fig:fitting_matrix}
\end{figure*}

As we are interested in the quasi-continuum, a specific area of the matrix where the nucleus is expected to follow a statistical behaviour was selected for further analysis, shown in Figure~\ref{fig:matrices} and Figure~\ref{fig:fitting_matrix}, and the $\gamma$-ray spectra were normalised such that each bin in the selected region would correspond to the decay probability $P(E_{x}, E_{\gamma})$, for a $\gamma$-ray energy of $E_{\gamma}$ from an excited state with energy $E_{x}$, as $\sum_{E_{\gamma}}P(E_{x}, E_{\gamma})=1$. Due to the carbon backing of the target, a significant background from the first excited state in ${}^{12}$C was present in the data. However, despite the excitation energy of this state at 4.44~MeV, due to the different reaction kinematics, the background appears at a reconstructed excitation energy of $\sim 7$~MeV. Thus, we have limited the high-energy range to below this background.
To extract the functional forms of the \ac{NLD} and the \ac{gSF}, a $\chi^{2}$ fitting procedure was carried simultaneously over all excitation energies and $\gamma$-ray energies in the selected region \cite{Guttormsen1987,Guttormsen1996,Schiller2000,Larsen2011}
\begin{equation}
P(E_{x},E_{\gamma})_{\mathrm{fit}}=\frac{\rho(E_{x}-E_{\gamma})\mathcal{T}(E_{\gamma})}{\sum_{E_{\gamma}}\rho(E_{x}-E_{\gamma})\mathcal{T}(E_{\gamma})},\label{eq:pexeg}
\end{equation}
where $\rho(E_{x}-E_{\gamma})=\rho(E_{\mathrm{f}})$ is the \ac{NLD} at the final state and $\mathcal{T}(E_{\gamma})$ are the transmission coefficients and they are related to the \ac{gSF}, $f_{XL}(E_{\gamma})$, is via the relation,
\begin{equation}
    f_{XL}(E_{\gamma}) = \frac{\mathcal{T}(E_{\gamma})}{2 \pi E_{\gamma}^{2L+1}},\label{eq:f2T}
\end{equation}
where $L$ is the multipolarity of the $\gamma$ ray and we assume $L=1$, corresponding to complete dipole domination, and $X$ correspond to either electric (E) or magnetic (M) transitions. While the total \ac{gSF} contains the sum of all $f_{XL}$, we can not explicitly separate the different multipolarity components in this data. However, measurements and theory generally support the assumption of complete dipole dominance. For example, using high-resolution forward angle inelastic proton scattering on ${}^{208}$Pb at the \ac{RCNP} in Japan, \cite{Bassauer2016} showed a $\sim$~10\% contribution to the \ac{gSF} from M1 around $S_{\mathrm{n}}$, which will not influence the evaluation in Equation~(\ref{eq:f2T}), and about 1\% contribution from E2 in the 10-13~MeV region, much higher in energy than the investigation presented here. The result of this fit is shown in Figure~\ref{fig:fitting_matrix}, together with two projections at selected energies to highlight the agreement between the fit and the experimental data. At this point, we evaluate the systematic uncertainties that can originate from the data processing as well as assumptions required for the method to work, from the reduced $\chi^{2}$ by taking the \ac{NDF} into account such that $\chi^{2}/\mathrm{NDF} = 1$, giving a contribution of $\sim 27$\% from the statistical and $\sim 73$\% from the systematic uncertainties to the total uncertainties. The methodological systematic uncertainties could, for example, include the validity of the Brink-Axel hypothesis that the \acp{gSF} are independent of $E_{\mathrm{x}}$ and, thus, that a unique \ac{gSF} will fit all the data. Here, we include these uncertainties and the statistical uncertainties in the analysis. We note that around $E_{\gamma}\approx 1.7$~MeV, there is an oversubtraction originating from the first-generation procedure, likely originating from a small ${}^{27}$Al background. This over-subtraction will propagate into the \ac{gSF} evaluation as an artificial dip. However, as the \ac{NLD} contains the sum over all possible energies, it will not introduce any artificial features. As this ${}^{27}$Al background appears in the raw matrix and, thus, has a correct reaction-kinematic reconstruction, it must originate from the target area, most likely from the target frame. While the 1.7~MeV transition is the strongest, a couple of weaker discrete transitions have been identified in Figure~\ref{fig:matrices}. Due to the low level density in ${}^{27}$Al, no statistical transitions are expected in the energy range of interest; thus, the contamination can be subtracted. The impact of this aluminium contamination is well below the statistical uncertainties in significance, even without the subtraction of the ${}^{27}$Al contribution, as shown in Figure~\ref{fig:matrices}. The ${}^{27}$Al contribution most prominently appears as an additional peak with an energy of 2.7~MeV and a more significant over-subtraction at 1.7~MeV.

\begin{figure*}[ht!]
\begin{center}
\includegraphics[width=0.49\textwidth]{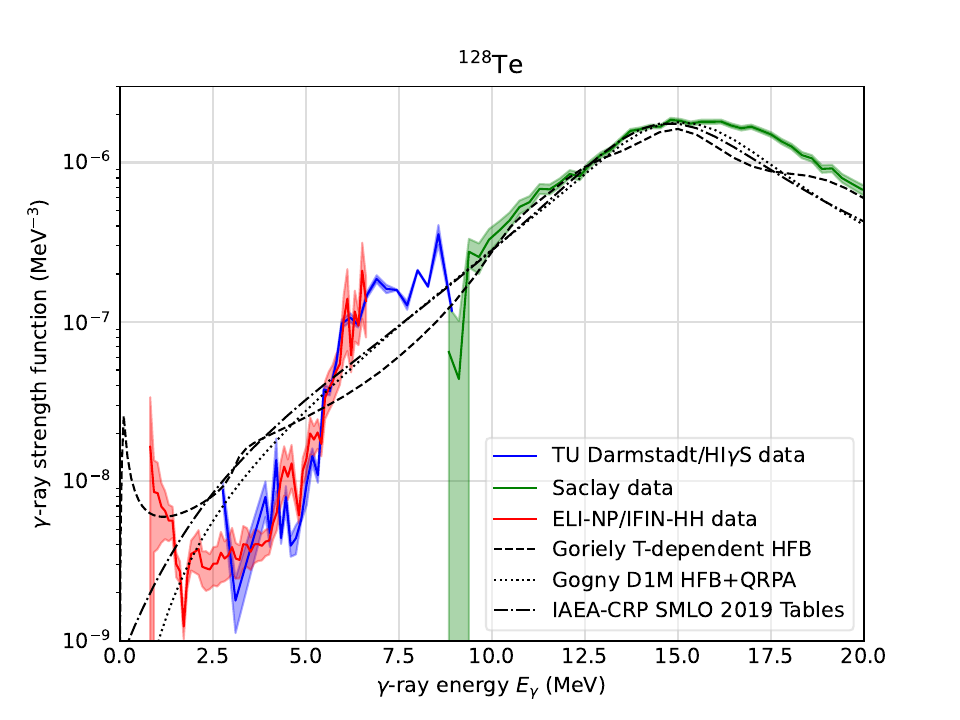}
\includegraphics[width=0.49\textwidth]{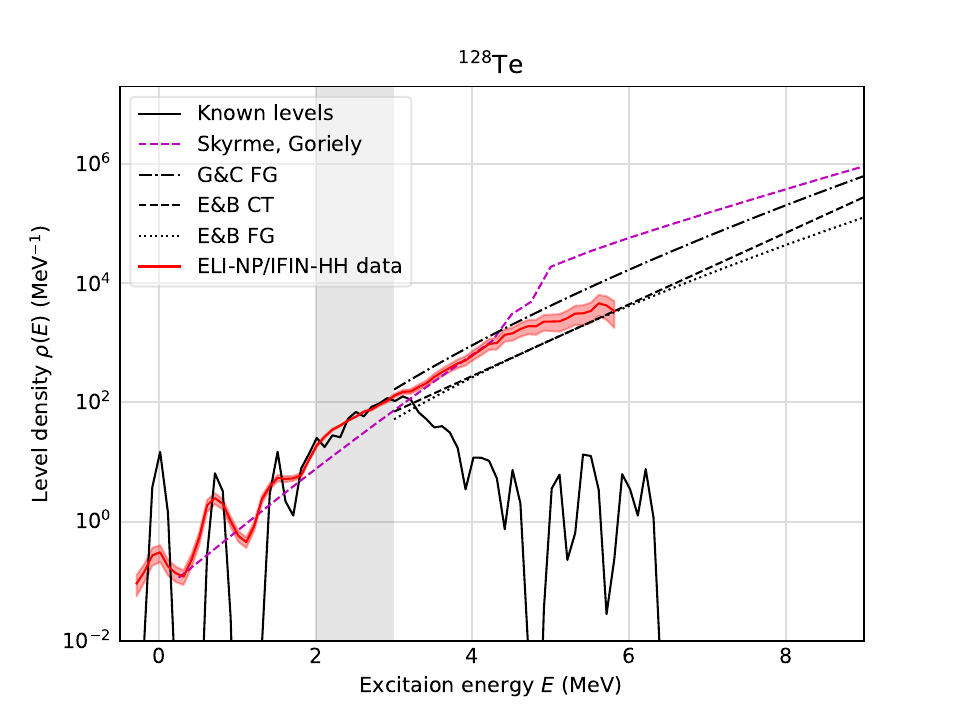}
\end{center}
\caption{(Left) Experimental $\gamma$-ray strength functions of ${}^{128}$Te obtained from this work (ELI-NP/IFIN-HH), the photoabsorption cross-section from Reference~\cite{Isaak2021} (TU Darmstadt/HI$\gamma$S), and $(\gamma,\mathrm{n})$ cross-section data from Reference~\cite{Lepretre1976} (Saclay). For comparison, three typical parametrisations, as implemented in the \talys\ code, are shown as solid lines for electric dipole strength and dashed lines for magnetic dipole strength. (Right) Nuclear level density of ${}^{128}$Te. The shaded region shows the range for fitting to known spectroscopic states. The transparent bands correspond to the statistical and systematic uncertainties. Three phenomenological parametrisations discussed in the text and one microscopic model implemented in \talys\ are shown for comparison.}\label{fig:te128_gsfnld}
\end{figure*}

It is well known that the Oslo method only gives the functional form of the \acp{NLD} and \acp{gSF}, while there are an infinite number of possible $\mathcal{T}(E_{\gamma})$ and $\rho(E_{x}-E_{\gamma})$ that can satisfy the fit in Equation~(\ref{eq:pexeg}). The relation between these possible $\mathcal{T}(E_{\gamma})$ and $\rho(E_{x}-E_{\gamma})$ is given by three parameters, $A$, $B$, and $\alpha$, as \cite{Guttormsen1987,Guttormsen1996,Schiller2000,Larsen2011},
\begin{eqnarray}
    \tilde{\rho}(E_{x}-E_{\gamma}) = A \rho(E_{x}-E_{\gamma})\exp[\alpha(E_{x}-E_{\gamma})],\label{eq:rho}\\
    \tilde{\mathcal{T}}(E_{\gamma}) = B \mathcal{T}(E_{\gamma})\exp(\alpha E_{\gamma})\label{eq:tau}.
\end{eqnarray}
By extracting $P(E_{\gamma}, E_{x})$ from the measured results, we have obtained one pair of solutions for $\rho(E_{x})$ and $\mathcal{T}(E_{\gamma})$. The normalisation procedure for obtaining the physical solution depends on the complementary information available for each case. In the regular Oslo method, the average neutron resonance spacing, $D_{0}$, is often used to obtain the \ac{NLD} at the neutron threshold, $\rho(S_{\mathrm{n}})$, to fix the slope of the \ac{NLD}, followed by fixing the absolute value of the \ac{gSF} to the average neutron resonance $\gamma$-ray decay width, $\langle\Gamma_{\gamma}\rangle$. In cases when the $D_{0}$ parameter is not available, other normalisation methods have been developed. The Shape method \cite{Wiedeking2021}, for example, have been developed to begin the normalisation procedure by fixing the shape of the \ac{gSF} through decay to discrete states and use the constrained \ac{gSF} to extract the \ac{NLD}. Here, we explore another option to constrain the \ac{gSF} for evaluating the \ac{NLD}. By normalising the $\mathcal{T}(E_{\gamma})$ solution to the photoabsorption cross-section \cite{Isaak2021}, we can extract the parameters $B$, corresponding to the absolute values, and $\alpha$, corresponding to the slope, from equation (\ref{eq:tau}), shown in Figure~\ref{fig:te128_gsfnld}. 
Note that we only include the photoabsorption cross-section from elastic and inelastic photon scattering from Reference~\cite{Isaak2021} and not the photoneutron data from Reference~\cite{Lepretre1976} in the minimisation. However, both data sets are shown in Figure~\ref{fig:te128_gsfnld} for reference. The reason for excluding the photoneutron data is the known issue with systematic discrepancies between old photoneutron measurements between different facilities \cite{Kawano2020}, which would propagate additional uncertainties into our results.
The fit to the data was done using a $\chi^{2}$ minimization on the \ac{gSF} as,
\begin{equation}
    \chi^{2} = \sum_{i} \frac{({f_{(\gamma,\gamma'),i}}-{\tilde{f}_{i})}^{2}}{\sigma_{{f_{(\gamma,\gamma'),i}}}^{2}+\sigma_{{\tilde{f}_{i}}}^{2}},
\end{equation}
where $f_{(\gamma,\gamma')}$ is the \ac{gSF} from photon scattering data and $\tilde{f}$ is the \ac{gSF} from the combination of Equations~(\ref{eq:f2T}) and (\ref{eq:tau}). The region included for the $\chi^{2}$ minimisation here was the full energy overlap region between our data and Reference~\cite{Isaak2021} giving a relative uncertainty of $\sigma_{\alpha}/\alpha=3.4$\% for a reduced $\chi^{2}/\mathrm{NDF} = 2.65$ when corrected for the number of degrees of freedom (NDF). With the slope known, we can fix the absolute value of the \ac{NLD}, parameter $A$, from the complete spectroscopy of known states in the energy range 2-3~MeV \cite{Elekes2015}. This range was chosen to have a lower energy limit above where several states start contributing to the \ac{NLD}, but below the energy limit where the experimental information of the number of excited states is incomplete. The latter was chosen according to the recommendations from \ac{RIPL}-3 \cite{Capote2009}, where the recommendation of the number of levels in a complete level scheme is 110, corresponding to 3.345~MeV. The typical uncertainty in the \ac{NLD} at $E_{x}>3$~MeV before normalization was $\sigma_{\rho(E_{x})}/\rho(E_{x}) \approx 13$\%. The contribution to the systematic uncertainty from the fitting of $\alpha$ is negligible close to $E_{x}=3$~MeV while increasing linearly for the extrapolation along $E_{x}$ for a total uncertainty of $\sigma_{\rho(E_{x})}/\rho(E_{x}) \approx 50$\% at the highest energy, $E_{x}=5.8$~MeV. The resulting \ac{NLD} is shown in Figure~\ref{fig:te128_gsfnld} and listed in Table~\ref{tab:expdata}, under the assumption that for at the \ac{gSF} is independent of the initial spin, and only depends on the $\gamma$-ray energy and electromagnetic multipolarity in the statistical region, such that the \ac{gSF} from the photoabsorption cross-section, mainly populating a very sharp spin ($J$) and parity ($P$) distribution with $J^{P} \approx 1^{\pm}$, is equivalent to the \ac{gSF} from higher spin states populated in inelastic $(\mathrm{p},\mathrm{p} '\gamma)$ scattering.

\begin{table}
\caption{\label{tab:expdata}Experimental results on the nuclear level density (NLD) in ${}^{128}$Te.}
\footnotesize
\begin{tabular}{@{}lclclclc}
\br
Energy & NLD & Energy & NLD & Energy & NLD & Energy & NLD\\
(MeV) & (MeV$^{-1}$) & (MeV) & ($\times 10$ MeV$^{-1}$) & (MeV) & ($\times 10^{2}$ MeV$^{-1}$) & (MeV) & ($\times 10^{3}$ MeV$^{-1}$)\\
\mr
-0.2875	&	0.09(4)	&	1.3125	&	0.238(35)	&	2.9125	&	1.07(6)	&	4.5125	&	1.44(35)	\\
-0.1875	&	0.15(5)	&	1.4125	&	0.39(5)	&	3.0125	&	1.32(9)	&	4.6125	&	1.7(4)	\\
-0.0875	&	0.27(9)	&	1.5125	&	0.54(7)	&	3.1125	&	1.50(12)	&	4.7125	&	1.9(5)	\\
0.0125	&	0.31(10)	&	1.6125	&	0.52(6)	&	3.2125	&	1.53(14)	&	4.8125	&	1.9(5)	\\
0.1125	&	0.18(5)	&	1.7125	&	0.53(5)	&	3.3125	&	1.80(19)	&	4.9125	&	2.3(7)	\\
0.2125	&	0.14(4)	&	1.8125	&	0.60(5)	&	3.4125	&	2.08(24)	&	5.0125	&	2.3(7)	\\
0.3125	&	0.119(33)	&	1.9125	&	1.09(8)	&	3.5125	&	2.67(35)	&	5.1125	&	2.3(7)	\\
0.4125	&	0.23(6)	&	2.0125	&	1.89(11)	&	3.6125	&	3.2(5)	&	5.2125	&	2.6(9)	\\
0.5125	&	0.54(13)	&	2.1125	&	2.64(12)	&	3.7125	&	3.8(6)	&	5.3125	&	3.1(11)	\\
0.6125	&	1.9(4)	&	2.2125	&	3.49(12)	&	3.8125	&	4.5(8)	&	5.4125	&	3.1(11)	\\
0.7125	&	2.5(5)	&	2.3125	&	4.08(11)	&	3.9125	&	5.0(9)	&	5.5125	&	3.4(13)	\\
0.8125	&	2.0(4)	&	2.4125	&	4.96(9)	&	4.0125	&	6.3(12)	&	5.6125	&	4.6(17)	\\
0.9125	&	1.02(20)	&	2.5125	&	5.76(10)	&	4.1125	&	7.7(15)	&	5.7125	&	4.2(18)	\\
1.0125	&	0.57(12)	&	2.6125	&	6.95(17)	&	4.2125	&	9.6(20)	&	5.8125	&	3.4(16)	\\
1.1125	&	0.45(9)	&	2.7125	&	7.59(30)	&	4.3125	&	9.9(22)	&		&		\\
1.2125	&	0.84(15)	&	2.8125	&	9.1(4)	&	4.4125	&	13.6(32)	&		&		\\

\br
\end{tabular}\\
\end{table}
\normalsize

\begin{figure*}[ht!]
\begin{center}
\includegraphics[width=0.49\textwidth]{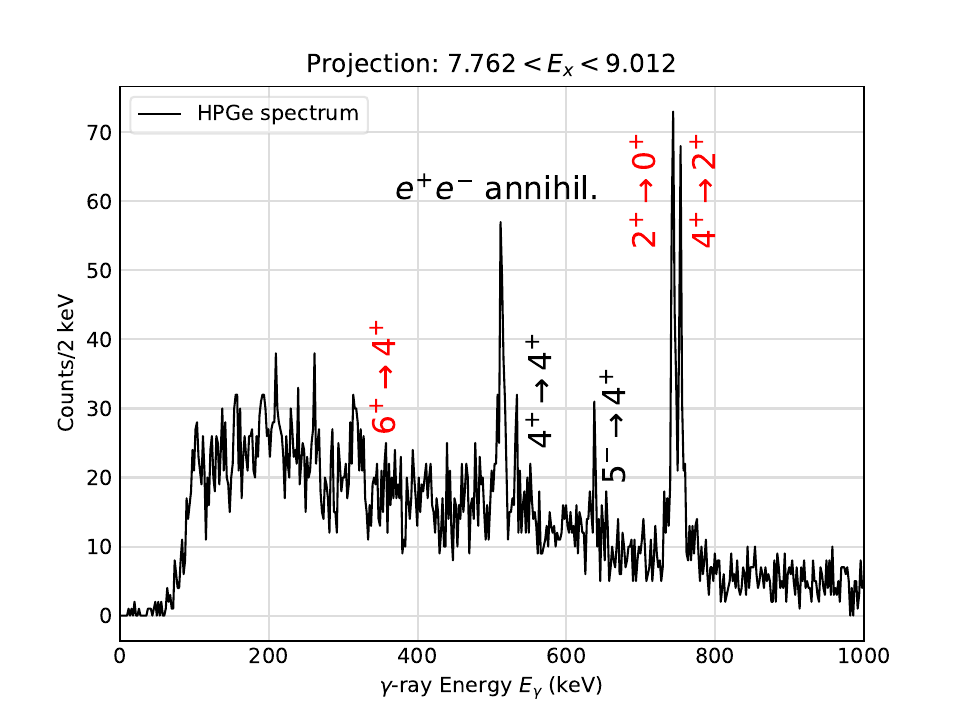}
\includegraphics[width=0.49\textwidth]{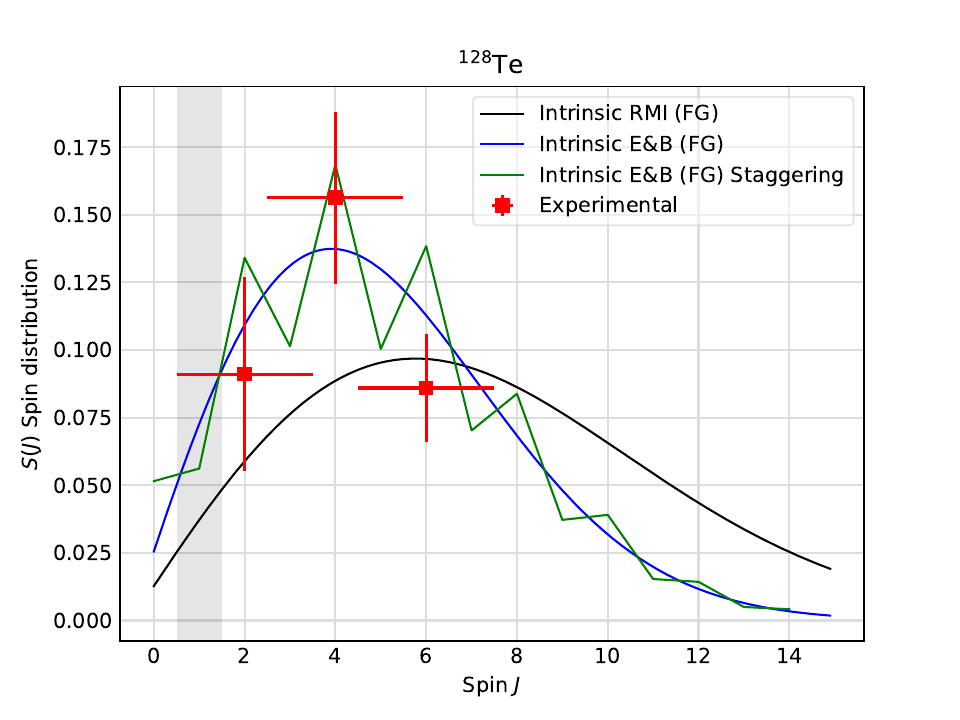}
\end{center}
\caption{(Left) HPGe spectrum showing the low-lying transitions in the yrast band, highlighted in red, from cascade $\gamma$ rays, and identified non-yrast transitions, with an excitation energy gate around the neutron separation threshold. (Right) Intrinsic nuclear spin distribution at the neutron threshold from three models shown with solid lines: the Rigid Moment of Inertia and the Fermi Gas with and without empirical odd-even staggering. The estimate of the experimental spin distribution from the HPGe detectors, consisting of the intrinsic spin distribution folded by the relative population cross-section, is shown as red points for comparison.\label{fig:spindist}}
\end{figure*}

The general features of the experimental spin distribution can be evaluated from the \ac{HPGe} detectors. For this purpose, an example of a \ac{HPGe} spectrum with an excitation energy gate between $7.762 < E_{x} < 9.012$~MeV is shown in Figure~\ref{fig:spindist}, showing the characteristic peaks associated with the $J^{\pi} = 6^{+} \to 4^{+}$, $J^{\pi} = 4^{+} \to 2^{+}$, and $J^{\pi} = 2^{+} \to 0^{+}$ transitions.
We can estimate the spin distribution $S(J)$ at spin $J$, from the efficiency corrected intensity of the $\gamma$-ray transitions, $I_{\gamma}$ populating and depopulating states with spin $J$ \cite{Guttormsen2022},
\begin{equation}
    S(J) = I_{\gamma}(J \to J-2) - I_{\gamma}(J+2 \to J).
\end{equation}
While this method gives stronger constraints for well-deformed nuclei with a clear population of high-spin values, it can still add some constraints to the spin range populated in this work. The subtracted and normalised spin distributions are shown in Figure~\ref{fig:spindist} together with three theoretical models based on
\begin{equation}
    S(J) = \frac{2J+1}{2 \sigma^{2}} \exp \left[ -\frac{\left( J+1/2 \right)^{2}}{2 \sigma^{2}} \right] \label{eq:sigma}.
\end{equation}
Figure~\ref{fig:spindist} shows the distribution for two values of $\sigma$, the Rigid Moment of Inertia (RMI) with $\sigma_{\mathrm{RMI}} = 6.263$ and the Fermi Gas model from von Egidy and Bucurescu \cite{vonEgidy2009} with $\sigma_{\mathrm{E\&B(FG)}} = 4.414$. The third model is the Fermi Gas model from von Egidy and Bucurescu with an additional staggering term added \cite{vonEgidy2008} as $S'(J) = S(J)(1+x)$ where $x=1.02$ for $J=0$, $x=0.227$ for even $J$ and $J=-0.227$ for odd $J$. It is important here to point out that these models provide us with the predicted intrinsic spin distribution of the statistical states, while the experimental data only give an estimate of the experimental spin distribution, meaning the intrinsic spin distribution is folded by the spin dependency of the population cross-section. The most extreme case in this work is the photonuclear data where almost the entire cross-section is in $J=1$, but also for the $(\mathrm{p},\mathrm{p}'\gamma)$ data, the spin-dependency of the population cross-section is expected to narrow down the experimental spin distribution \cite{Guttormsen2022}.

\section{Discussion}

The overall agreement between the shape of the \ac{gSF} from the photoabsorption cross-section \cite{Isaak2021} is shown in Figure~\ref{fig:te128_gsfnld}. Note that since the photoabsorption cross-section was used to normalise the absolute magnitude and the total slope of the \ac{gSF}, only the relative shapes can be used to evaluate the consistency. For comparison, we also include the following microscopic calculations and parametrisations of the \acp{gSF} as implemented in the \talys\ \cite{Koning2008,Koning2012} code: the temperature-dependent \ac{HFB} by Goriely \cite{Hilaire2012},  the \ac{HFB} and \ac{QRPA} calculations with the Gogny D1M interaction \cite{Goriely2018}, and the \ac{SMLO} function from the recommendations of the \ac{IAEA} \ac{CRP} on \acp{gSF} \cite{Kawano2020,Goriely2019b}.

On the \ac{NLD} side, the procedure described in Section~\ref{sec:analysis} will provide us with the \ac{NLD} directly from experimental data without the need to invoke model-dependent extrapolations to the neutron separation energy, typically via the constant temperature or the Fermi gas model. Thus, comparing the experimental data presented here to these two common extrapolations would be interesting. The two models are simple phenomenological two-parameter models that are based on the observation of empirical nuclear properties. Here, we will use the formulation of these models presented in Reference~\cite{vonEgidy1988}.

The constant temperature model \cite{Gilbert1965} is based on the observation that the number of nuclear levels typically increases with a simple exponential behaviour with increasing excitation energy, implying the nucleus stays at a constant temperature, $T$, in the applicable energy range. This description is justified by the increased breaking of Cooper pairs with increasing excitation energy, keeping the mean energy per nucleon constant. Thus, the \ac{NLD} as a function of $E_{\mathrm{x}}$ and $J$, in the constant temperature approximation, $\rho_{\mathrm{CT}}(E_{\mathrm{x}},J)$, can be expressed as 
\begin{equation}
    \rho_{\mathrm{CT}}(E_{\mathrm{x}},J) = f(J) \frac{1}{T} \exp\left(\frac{E_{\mathrm{x}}-E_{0}}{T}\right). \label{eq:ct}
\end{equation}
In addition to $T$, a back-shift parameter, $E_{0}$, has been introduced to adjust for the finite energy where the statistical properties dominate. 

The Fermi gas \ac{NLD} \cite{vonEgidy1988}, $\rho_{\mathrm{FG}}(E_{\mathrm{x}}, J)$, on the other hand, is justified considering the nucleus as a gas of non-interacting fermions within the nuclear volume and calculated the number of possibilities to distribute the excitation energy on these \cite{Bethe1936}, considering single-particle states,
\begin{equation}
    \rho_{\mathrm{FG}}(E_{\mathrm{x}},J) = f(J) \frac{\exp\left[2 a (E_{\mathrm{x}}-E_{1})\right]}{12 \sqrt{2} \sigma a^{1/4} (E_{\mathrm{x}}-E_{1})^{5/4}}. \label{eq:fg}
\end{equation}
This expression also consists of two main parameters, the level density parameter $a$ and the Fermi back-shift parameter $E_{1}$. In addition, the spin cut-off parameter, $\sigma$, is included explicitly in the expression (\ref{eq:fg}). In fact, in both equations (\ref{eq:ct}) and (\ref{eq:fg}), the spin distribution has been considered independent from $E_{x}$ and separated into the function $S(J)$ from Equation~(\ref{eq:sigma}). 
Note that this energy independence of $S(J)$ is an approximation. In an actual nucleus, there is a likely energy dependence of $\sigma$ and, therefore, also on $S(J)$.

To compare the experimental data with the phenomenological descriptions in equations (\ref{eq:ct}-\ref{eq:sigma}), we employed a selection of typical parametrisations of the data from Gilbert and Cameron \cite{Gilbert1965}, and von Egidy and Bucurescu \cite{vonEgidy2009}. The extraction of parameters for equations (\ref{eq:ct}-\ref{eq:sigma}) was performed using the ROBIN code version 1.92 \cite{GuttormsenUnpROBIN}. The obtained parameters are listed in Table~\ref{tab:robin}, and the corresponding \acp{NLD} are shown in Figure~\ref{fig:te128_gsfnld}.
\begin{table}
\caption{\label{tab:robin}Parameters for different parametrisations of the constant temperature (CT) and Fermi gas (FG) models by Gilbert and Cameron (G\&C) and von Egidy and Bucurescu (E\&B). The level density parameter is denoted $a$, the Fermi back-shift $E_{1}$, the nuclear temperature $T$, and the constant temperature back-shift $E_{0}$. We also list the calculated temperature $T(S_{\mathrm{n}})$, nuclear level density $\rho(S_{\mathrm{n}})$, and spin cut-off $\sigma(S_{\mathrm{n}})$, at the neutron threshold $S_{\mathrm{n}}$.}
\footnotesize
\begin{tabular}{@{}lccccccc}
\br
Model & $a$ & $E_{1}$ & $T$ & $E_{0}$ & $T(S_{\mathrm{n}})$ & $\rho(S_{\mathrm{n}})$ & $\sigma(S_{\mathrm{n}})$\\
 & MeV$^{-1}$ & MeV & MeV & MeV & MeV & MeV$^{-1}$ &\\
\mr
G\&C (FG) \cite{Gilbert1965} & $13.073$ & $0.680$ &  &  & 0.78723 & $0.40912 \times 10^{6}$ & 4.818\\
E\&B (CT) \cite{vonEgidy2009} &  &  & $0.724$ & $0.159$ & 0.724 & $0.20476 \times 10^{6}$ & 4.002\\
E\&B (FG) \cite{vonEgidy2009} & $11.738$ & $0.782$ &  &  & 0.82556 & $0.14 \times 10^{6}$ & 4.414\\
\br
\end{tabular}\\
\end{table}
\normalsize
In addition to the phenomenological models, the calculated results from the \ac{RIPL} calculated from the Skyrme force, as implemented in \talys\ 1.95, are also shown \cite{Goriely2001}.

\section{Conclusion and outlook}

We have extracted the nuclear level density of ${}^{128}$Te from a $(\mathrm{p},\mathrm{p} '\gamma)$ scattering experiment using the large-volume \labr\ and \cebr\ detectors from ELI-NP at the 9~MV tandem facilities at IFIN-HH. The decay data were normalised using the photoabsorption cross-section, which provides nuclear level densities without intrinsic model dependencies from the constant temperature or Fermi gas models. The nuclear level density follows in between the expectations from these two models, but we observe a clear divergence from the microscopic model based on the Skyrme force. Further work will include the in-depth study of the $\gamma$-ray strength functions from this data, in particular, to investigate how assumptions like the validity of the Brink-Axel hypothesis and the contribution from the spin distribution affect $\gamma$-ray strength-function results in both photonuclear data and charged particle data.
While the work presented here aimed at removing some of the model dependencies present in the analysis methods, there are still uncertainties present regarding, for example, the spin distribution that is difficult to resolve completely with current data and methods. Systematic investigations of pure $J^{\pi}=1^{-}$ \acp{NLD} with the future ELI-NP photon beams will further help pin down these uncertainties and the modelling of \acp{NLD}, partially addressing the remaining uncertainties or method dependent characteristics of \ac{NLD} data.

\section*{Acknowledgements}

PAS, AK, RB, MB, CCo, NMF, AGa, RL, CMi, AS, DAT, AT, GVT, and SU were supported by the ELI-RO program funded by the Institute of Atomic Physics, M\u{a}gurele, Romania, contract number ELI-RO/RDI/2024-002 (CIPHERS) and SA, DLB, SRB, RC, and TP ELI-RO/RDI/2024-007 (ELITE). The remaining authors from ELI-NP acknowledge the support of the Romanian Ministry of Research and Innovation under research contract PN~23~21~01~06. This project has received funding from the European Union's Horizon Europe Research and Innovation Programme under Grant Agreement No 101057511 (EURO-LABS). We would also like to acknowledge the International Atomic Energy Agency Coordinated Research Activity F41034 ``Updating and Improving Nuclear Level Densities for Applications''. This work was presented at the International Symposium on Nuclear Science (ISNS-24), Sofia 2024.

\section*{References}
\providecommand{\newblock}{}

\acrodef{AGATA}{Advanced GAmma Tracking Array}
\acrodef{ALBA}{African LaBr Array}
\acrodef{AMD}{antisymmetrized molecular dynamics}
\acrodef{BGO}{bismuth germanate}
\acrodef{BNC}{Bayonet Neill-Concelman}
\acrodef{BRIKEN}{Beta-delayed neutrons at RIKEN}
\acrodef{BRUSLIB}{BRUSsels Nuclear LIBrary}
\acrodef{CAKE}{Coincidence Array for K600 Experiment}
\acrodef{CAD}{computer-aided design}
\acrodef{CCB}{Centrum Cyklotronowe Bronowice}
\acrodef{CFD}{constant-fraction discriminator}
\acrodef{CLHEP}{Class Library for High Energy Physics}
\acrodef{CMB}{cosmic microwave-background}
\acrodef{CRP}{Coordinated Research Project}
\acrodef{DAQ}{data acquisition}
\acrodef{DC}{direct current}
\acrodef{DELILA}{Digital ELI List-mode Acquisition}
\acrodef{DPP}{Digital Pulse-Processing}
\acrodef{DPP-PHA}{Digital Pulse Processing for Pulse Height Analysis}
\acrodef{DPP-PSD}{Digital Pulse Processing for Charge Integration and Pulse Shape Discrimination}
\acrodef{DSSSD}{double-sided silicon strip detector}
\acrodef{D-sub}{D-subminiature}
\acrodef{E8}{Experimental Area 8}
\acrodef{E9}{Experimental Area 9}
\acrodef{ECL}{emitter-coupled logic}
\acrodef{EDF}{Energy Density Functional}
\acrodef{ELI}{Extreme Light Infrastructure}
\acrodef{ELIADE}{ELI Array of DEtectors}
\acrodef{ELI-BIC}{ELI Bragg ionization chamber}
\acrodef{ELIGANT}{ELI Gamma Above Neutron Threshold}
\acrodef{ELIGANT-GG}{ELIGANT Gamma Gamma}
\acrodef{ELIGANT-GN}{ELIGANT Gamma Neutron}
\acrodef{ELIGANT-TN}{ELIGANT Thermal Neutron}
\acrodef{ELI-NP}{Extreme Light Infrastructure -- Nuclear Physics}
\acrodef{ELISSA}{ELI Silicon Strip Array}
\acrodef{EWSR}{energy-weighted sum rule}
\acrodef{FADC}{flash ADC}
\acrodef{FASTER}{Fast Acquisition System for nuclEar Research}
\acrodef{FATIMA}{FAst TIming Array}
\acrodef{FFT}{fast Fourier transform}
\acrodef{FOM}{figure-of-merit}
\acrodef{FREYA}{Fission Reaction Event Yield Algorithm}
\acrodef{FWHM}{full width at half maximum}
\acrodef{GDR}{giant dipole-resonance}
\acrodef{GECO2020}{GEneral COntrol Software}
\acrodef{GLO}{Generalized Lorentzian}
\acrodef{GRAF}{Grand RAiden Forward}
\acrodef{GROOT}{\geant\ and ROOT Object-Oriented Toolkit}
\acrodef{gSF}[$\gamma$SF]{$\gamma$-ray strength function}
\acrodef{GSI}{Gesellschaft f\"{u}r Schwerionenforschung}
\acrodef{HDPE}{high-density polyethylene}
\acrodef{HF}{Hartree-Fock}
\acrodef{HFB}{Hartree-Fock-Bogolyubov}
\acrodef{HIgS}[HI$\gamma$S]{High-Intensity $\gamma$-ray Source}
\acrodef{HPGe}{high-purity germanium}
\acrodef{HPLS}{high-power laser system}
\acrodef{HV}{high voltage}
\acrodef{IAEA}{International Atomic Energy Agency}
\acrodef{IFIN-HH}{Horia Hulubei Institute for Physics and Nuclear Engineering}
\acrodef{iThemba LABS}{iThemba Laboratory for Accelerator Based Sciences}
\acrodef{JINR}{Joint Institute for Nuclear Research}
\acrodef{KMF}{Kadmenskii-Markushev-Furman}
\acrodef{KRATTA}{Krak\'{o}w Triple Telescope Array}
\acrodef{LANL}{Los Alamos National Laboratory}
\acrodef{LCS}{laser Compton backscattering}
\acrodef{LED}{leading-edge discriminator}
\acrodef{LEDR}{low-energy electric dipole response}
\acrodef{LLNL}{Lawrence Livermore National Laboratory}
\acrodef{MDA}{multipole decomposition analysis}
\acrodef{MDR}{magnetic dipole-resonance}
\acrodef{MCA}{multi-channel analyzer}
\acrodef{MCNP}{Monte Carlo N-Particle Transport Code}
\acrodef{MCX}{micro coaxial connector}
\acrodef{MIDAS}{Multi Instance Data Acquisition System}
\acrodef{MONSTER}{MOdular Neutron time-of-flight SpectromeTER}
\acrodef{MWDC}{Multi-Wire Drift Chambers}
\acrodef{NDF}{number of degrees of freedom}
\acrodef{NEDA}{NEutron Detector Array}
\acrodef{NIM}{Nuclear Instrumentation Module}
\acrodef{NLD}{nuclear level density}
\acrodefplural{NLD}[NLDs]{nuclear level densities}
\acrodef{NNDC}{National Nuclear Data Center}
\acrodef{NRF}{nuclear resonance fluorescence}
\acrodef{OCL}{Oslo Cyclotron Laboratory}
\acrodef{ODeSA}{Oak Ridge National Laboratory Deuterated Spectroscopic Array}
\acrodef{ORNL}{Oak Ridge National Laboratory}
\acrodef{OSCAR}{Oslo Scintillator Array}
\acrodef{PANDORA}{Photo-Absorption of Nuclei and Decay Observables for Reactions in Astrophysics}
\acrodef{PARIS}{Photon Array for Studies with Radioactive Ion and Stable Beams}
\acrodef{PCB}{printed circuit board}
\acrodef{PCIe}{Peripheral Component Interconnect Express}
\acrodef{PDR}{pygmy dipole resonance}
\acrodef{PLL}{phase-locked loop}
\acrodef{PMT}{photomultiplier tube}
\acrodef{PRD}{prompt-response distribution}
\acrodef{PSD}{pulse-shape discrimination}
\acrodef{PuBe}{plutonium-beryllium}
\acrodef{QED}{quantum electrodynamics}
\acrodef{QPM}{quasiparticle-phonon model}
\acrodef{QRPA}{Quasiparticle Random Phase Approximation}
\acrodef{RC}{resistor-capacitor}
\acrodef{RCNP}{Research Center for Nuclear Physics}
\acrodef{RF}{radio frequency}
\acrodef{RIPL}{Reference Input Parameter Library}
\acrodef{RMS}{root-mean-square}
\acrodef{ROSPHERE}{ROmanian array for SPectroscopy in HEavy ion REactions}
\acrodef{RQRPA}{Relativistic Quasiparticle Random Phase Approximation}
\acrodef{RQTBA}{Relativistic Quasiparticle Time-Blocking Approximation}
\acrodef{SHV}{safe high voltage}
\acrodef{SAKRA}{Si Array developed by Kyoto and osaka for Research into Alpha cluster states}
\acrodef{SMLO}{simplified version of the modified Lorentzian}
\acrodef{SORCERER}{SOlaR CElls for Reaction Experiments at ROSPHERE}
\acrodef{SSC}{Separated Sector Cyclotron}
\acrodef{TCP/IP}{Transmission Control Protocol/Internet Protocol}
\acrodef{TDR}{technical design report}
\acrodef{TRK}{Thomas-Reiche-Kuhn}
\acrodef{TOF}{time-of-flight}
\acrodef{TUD}[TU Darmstadt]{Technische Universität Darmstadt}
\acrodef{TUNL}{Triangle Universities Nuclear Laboratory}
\acrodef{UHECR}{ultra-high energy cosmic ray}
\acrodef{UPC}{Universitat Polit\`{e}cnica de Catalunya}
\acrodef{USB}{Universal Serial Bus}
\acrodef{VECC}{Variable Energy Cyclotron Centre}
\acrodef{VEGA}{Variable Energy Gamma-ray}
\acrodef{VME}{Versa Module Europa}

\end{document}